\documentclass[11pt]{iopart}
\usepackage{graphicx}
\usepackage{amssymb}
\usepackage{color}

\begin{document}
\title{Data Quality Studies of Enhanced Interferometric Gravitational Wave Detectors}
\footnote{Presented at the 9th Edoardo Amaldi Conference on Gravitational Waves, Cardiff, UK}
\date{\today}
\author{Jessica McIver$^1$ for the LIGO Scientific Collaboration and the Virgo Collaboration}
\address{$^1$  Department of Physics, University of Massachusetts, Amherst, MA 01003}

\ead{\mailto{jlmciver@physics.umass.edu}}

\begin{abstract}
Data quality assessment plays an essential role in the quest to detect gravitational wave signals in data from the LIGO and Virgo
interferometric gravitational wave detectors. Interferometer data contains a high rate of noise transients from the environment,
the detector hardware, and the detector control systems. These transients severely limit the statistical significance of gravitational
wave candidates of short duration and/or poorly modeled waveforms. This paper describes the data quality studies
that have been performed in recent LIGO and Virgo observing runs to mitigate the impact of transient detector
artifacts on the gravitational wave searches. 

\end{abstract}


%%%##\footnote{Presented at the 9th Edoardo Amaldi Conference on Gravitational Waves, Cardiff, UK}

%\titlepage

%~~~~~~~~~~~~~~~~~~~~~~~~~~~~~~~~~~~~~~~~~~~~~~~~~~
\section{Laser Interferometers and Data Quality}

Laser interferometer gravitational wave detectors are designed to measure small variations in the relative separation of mirrors at the ends of two perpendicular arms, corresponding to gravitational-wave (GW) strain caused by astronomical sources~\cite{Abramovici:1992}. In the course of these studies, a global network of {\em enhanced} GW detectors, including the Laser Interferometer Gravitational wave Observatory (LIGO) detectors in Hanford, WA and Livingston, LA each with 4 km arms, and the Virgo detector, with 3 km arms, in Cascina, Italy, were jointly taking data. The enhanced era science runs provided an opportunity to test ground-breaking detector technology and improve the detectors' sensitivity in preparation for the {\em advanced} detector era~\cite{Smith:2009bx}. During LIGO's science run 6 and Virgo's science runs 2 and 3, the detectors were capable of measuring differences in length up to one part in $10^{21}$ \cite{Abbott:2007kva, Accadia2011}, which is roughly the strain expected from transient gravitational wave sources such as coalescing systems of binary black holes. 

Such GW signals can be easily masked or mimicked by transient noise in the detector outputs. A wide variety of transient noise sources have been observed in LIGO and Virgo data. The LIGO and Virgo instruments are equipped with thousands of {\em Auxiliary channels}, including sensors monitoring environmental and instrumental variables, thar are used to identifying sources of non-astrophysical noise artefacts. Environmental auxiliary channels monitor variables such as seismic motion, changes in local magnetic field, and acoustics at key locations such as the enclosures that contain the detectors' most sensitive equipment. Similarly, instrumental auxiliary channels monitor systems such as laser beam alignment, laser stabilization, and mirror suspension controls. 

Noise transients play an especially important role in searches for short-duration (in the detectors' bands 40-10000Hz) or unmodeled gravitational waves produced by compact binary coalescence (CBC) and burst sources. CBC searches use a matched-filter analysis, which requires candidate events to match a template of known waveforms \cite{S5inspiral}. The higher the total mass of the systems of black holes and/or neutron stars, the shorter the duration of the signal in our detectors, and the more difficult it is to distinguish from a noise artefact. The search for bursts is an unmodeled and largely unconstrained analysis, and the multi-detector burst analysis pipelines look for coincident excess power in the signals and for coherence between the detectors \cite{S5burstAllSky, S5VSR1burst}. To improve the astrophysical range and detection likelihood of our astrophysical searches, we perform studies of the LIGO and Virgo data quality and use the information gained to mitigate transients in the GW channels of the detectors.

%--------------------------------------------------------------------
\section{Introduction to Noise Transients: Characteristics and Impact}

Noise transients in the GW data channel are characterized by their loudness or signal-to-noise ratio (SNR), central frequency, and spectrogram morphology. These characteristics are used to identify {\em populations} of noise transients that share similar traits. 

\begin{figure}
\centering
\begin{tabular} {cc}
\includegraphics[width=0.52\textwidth]{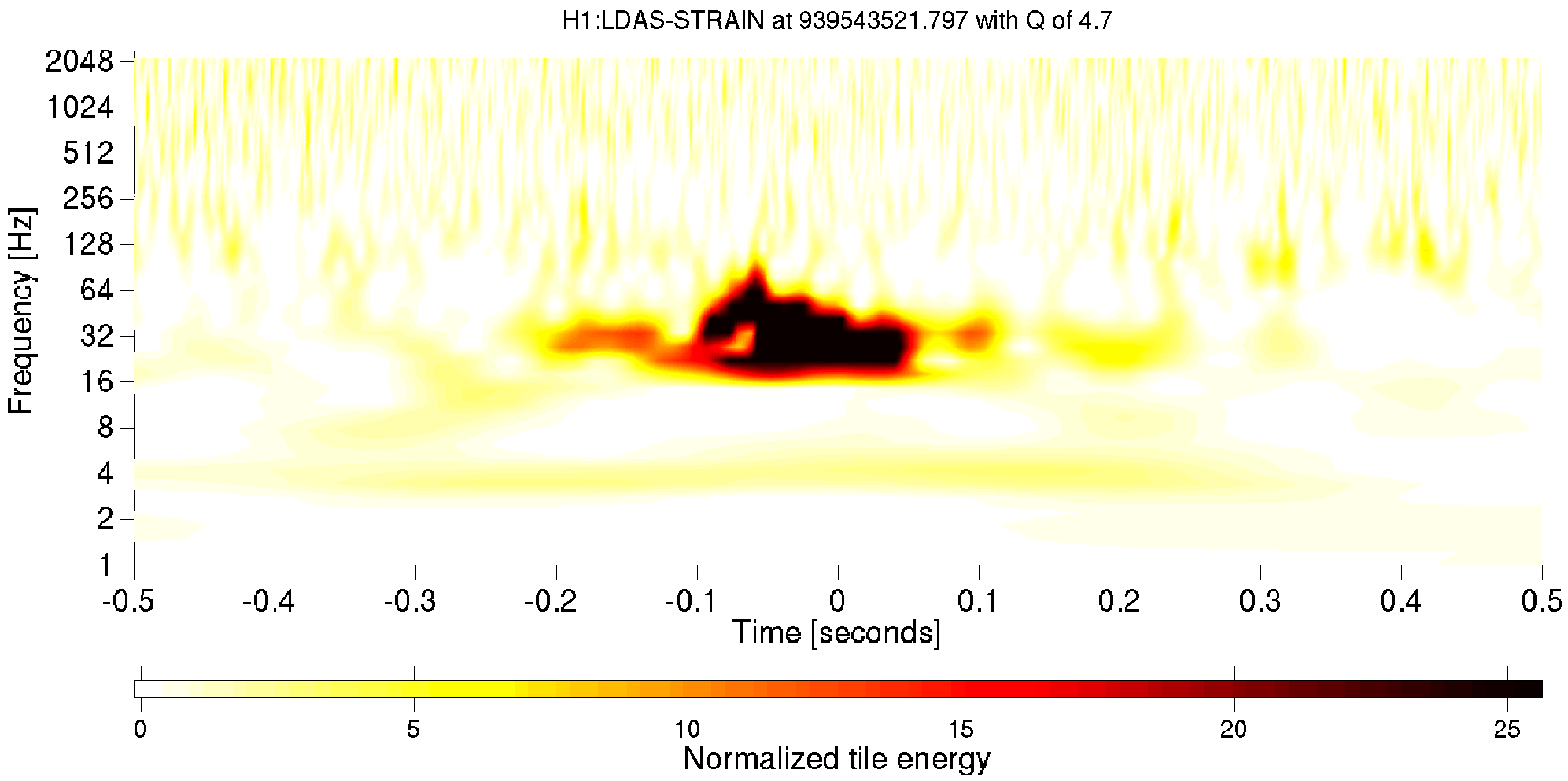}  &
\includegraphics[width=0.48\textwidth]{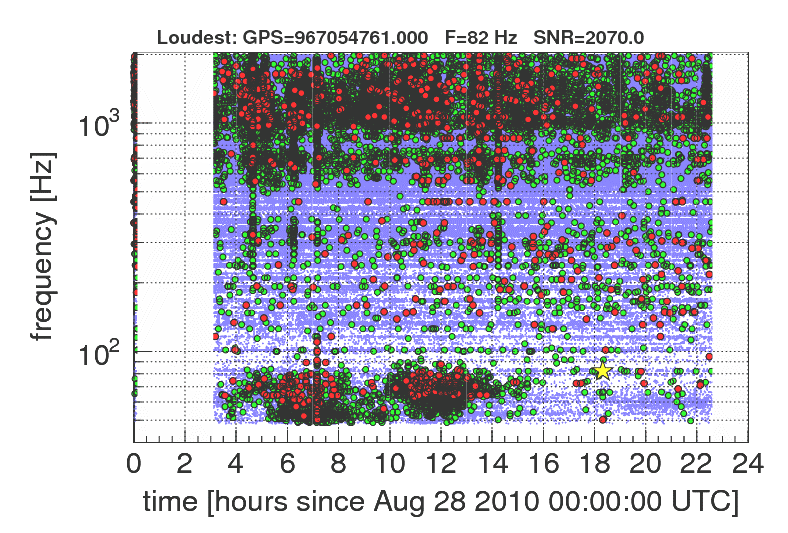} \\

(a) Spectrogram of a noise transient & (b) Noise transient time series\\
\end{tabular}
\caption{\small{\textit{(a) An example spectrogram depicting a noise transient in the GW data channel of the Hanford LIGO detector during the 6th LIGO science run, as seen through one of the single detector search algorithms, Omega \cite{Chatterji:2004, chatterji05:_ligo}. Here color represents energy (with white being the lowest and red the highest. (b) An example time-frequency plot of noise transients in the Virgo GW data channel as seen by Omega. The smaller blue points represent lower SNR transients ($>$5), and green and red represent progressively higher ranges of SNRs ($>$10, $>$20). 
}}}
\label{f:Omegascanglitchgram}
\end{figure}

Figure~\ref{f:Omegascanglitchgram}a shows a time-frequency spectrogram of one noise transient identified in data from the LIGO Hanford detector. This spectrogram was produced using Omega, a low-latency burst search algorithm~\cite{chatterji05:_ligo}. For each transient (noise or GW) it detects, Omega also outputs a set of numbers that parametrize the transient, including its central time, frequency, and SNR. This is what we call a ``trigger''. The transient shown in figure~\ref{f:Omegascanglitchgram}a corresponds to one of the red dots in the time-frequency plot of all of the Omega triggers for one day at Hanford shown in figure~\ref{f:Omegascanglitchgram}b. These plots indicate that gravitational-wave data contains many artefacts.  

\begin{figure}
\centering
\begin{tabular} {cc}
\includegraphics[width=0.5\textwidth]{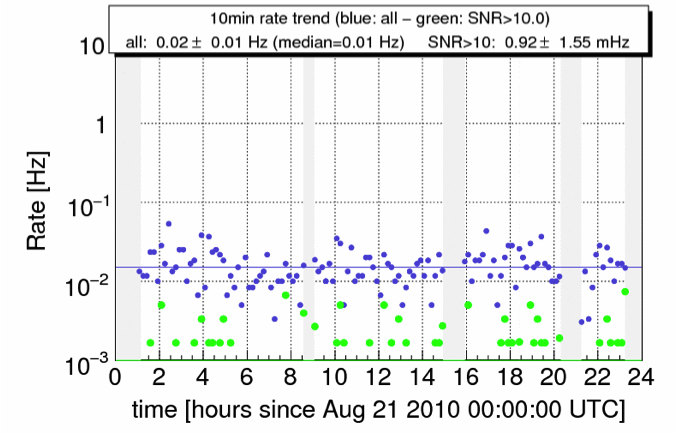}  &
\includegraphics[width=0.5\textwidth]{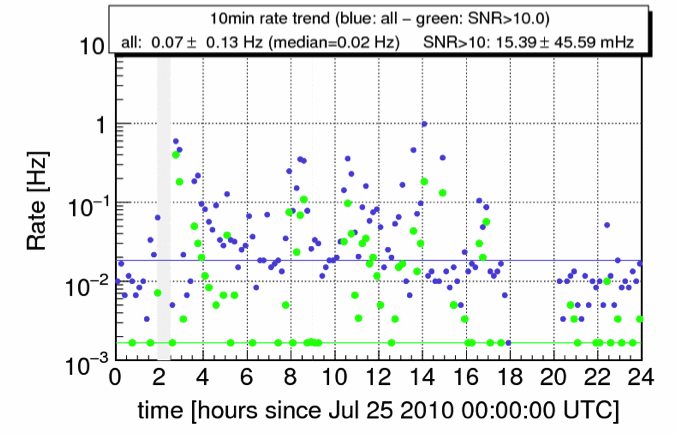} \\
\includegraphics[width=0.5\textwidth]{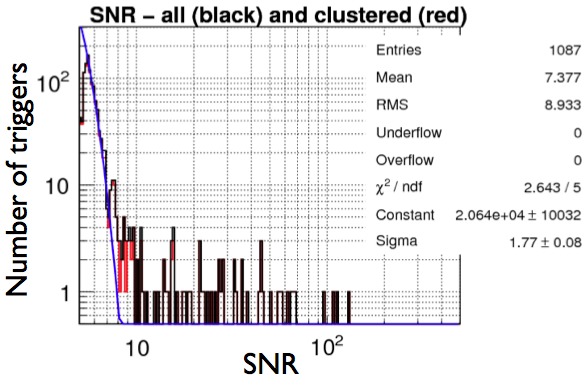}  &
\includegraphics[width=0.5\textwidth]{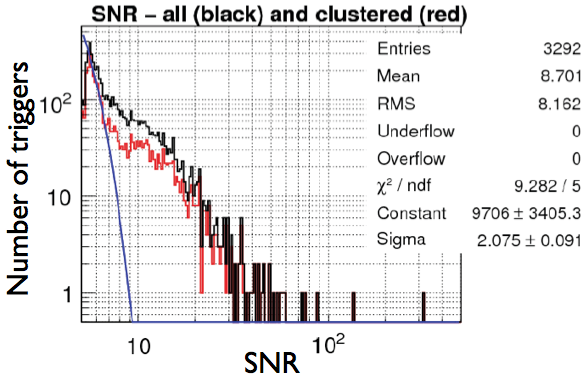} \\
(a) ``good day'' & (b) ``bad'' day\\
\end{tabular}
\caption{\small{\textit{Rate time series (above) and histograms (below) of Omega triggers from the LIGO Hanford detector output. Blue dots represent triggers of all SNRs and green dots represent
triggers of SNR $>$10. The lower plots compare the triggers SNR distribution (black) and "clustered" triggers of SNR $>$10 (red) to the expected Gaussian distribution, in blue. On a ``good" day, the trigger distribution is much closer to Gaussian (shown as the smooth blue curve that crosses the axis at an SNR or 8-9). The ``good" and ``bad" days were each taken from different times during S6, but have comparable detector livetime.
}}}
\label{f:goodbad}
\end{figure}

The panels in figure~\ref{f:goodbad} provide some context for the loudness and the rate of such triggers seen during LIGO's S6 run. Column ``a" shows what is considered a ``good" day (low-to-normal rate of noise transients), and column ``b" shows a ``bad" day (high rate of noise transients). When the interferometers are performing optimally, one could expect to see roughly one noise transient with an SNR $>$10 every couple of minutes, which is still significantly higher than the predicted Gaussian noise curve shown in blue in the rate histograms of figure~\ref{f:goodbad}.  How often ``bad" days occur depends largely on various external factors and potential sources
noise transients in the detectors, including bad weather, nearby train, truck, or air traffic, fluctuating magnetic fields around equipment, seismic noise, digital signal processing
artefacts, or malfunctioning equipment.  

The first line of defense against a noise transient masquerading as a potential GW signal is requiring coincidence or coherence between interferometers in the global network. However, noise transients do occur frequently and loudly enough to affect the searches, even when coherence is required. In figure 3, a histogram of triggers (transient noise events) as seen by a Coherent WaveBurst, an unmodeled search algorithm requiring coincidence and coherence between the detectors \cite{cWB}, is overlaid on a histogram of triggers seen by the single interferometer unmodeled search algorithm Omega. This illustrates the reduced effect of noise transients on the GW analyses by using a global network of detectors. It also gives an indication of the range of SNR of single-detector triggers that has the most impact on coherent unmodeled searches; in this case, a solid population of triggers with an SNR $<$20 and a few triggers with an SNR up to 200. 

Each search is affected differently by noise transients. For example, the CBC search, based on templated match-filter, is primarily affected by loud noise transients or noise transients that mimic the template forms, while the unmodeled search requiring coincidence is more affected by very frequent noise transients, which are more likely to be coincident. It is the responsibility of the LIGO and Virgo detector characterization groups to limit the impact of these noise transients as much as possible, using methods outlined below. 

\begin{figure}
\centering
\includegraphics[width=0.6\textwidth]{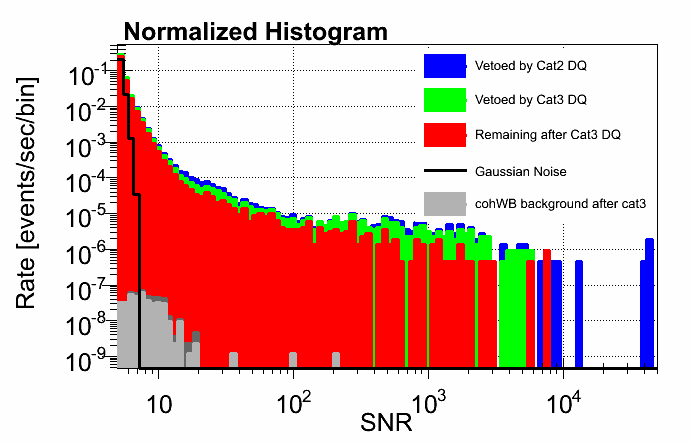}
\caption{\small{\textit{This histogram shows the SNR distribution of the Omega triggers for LIGO Hanford during S6A. The different color indicate the number of triggers before and after the application of (category 2 and 3) data quality flags. An additional black line indicates the trigger rate expected from Gaussian noise. The bottom set of grey triggers represent the search background (produced using time-slide analysis) for an unmodeled burst analysis that requires coincidence and coherence between detectors~\cite{cWB}. Further improvement can be obtained by requiring candidate triggers to match a modeled waveform~\cite{S5inspiral,S5burstAllSky,S5VSR1burst}.}}}
\label{f:cWBhist}
\end{figure}

%~~~~~~~~~~~~~~~~~~~~~~~~~~~~~~~~~~~~~~~~~~~~~~~~~~
\section{Methods Used to Abate Noise Transients}

Noise transients are generally dealt with in one of two ways. Ideally, investigations lead to clues that allow us to track down the source of the noise and/or the coupling mechanism to the GW data channel, and fix the problem so that it no longer pollutes the data. The second option is to cut the ``bad" parts out of the data, using data quality flags. This is less desirable since it reduces the overall time available for the analysis, and may not clean all of the associated artefacts. 

\subsection {Identification of noise transient sources}
The detector characterization group focuses on populations of noise transients that are troublesome for astrophysical searches, generally far outside Gaussian noise behavior. Sources of noise transients can vary wildly from external/environmental factors such as ocean waves, to equipment malfunction such as loose wire contacts. We can use the characteristics of the noise transients as clues to help track down the corresponding noise source. Various examples of environmental noise sources coupling to the GW data channel via instrumental imperfection or malfunctioning were observed during the scientific runs. For example, scattered light due to flaws in optical components is a mechanism through which seismic noise coupled into the data. 

The data quality teams routinely check noise transients for coincidence with the detector's auxiliary channels. Algorithms such as Hierarchical Veto~\cite{Hveto:2011} and Used Percentage Veto~\cite{UPV:2010}, produce statistical results ranking the coupling of auxiliary channels with the GW data channel on a daily and weekly basis. These suggest which channels or families of channels are the most closely associated with noise transients during that time. This information often leads to noise mitigation or data quality flag development.

%~~~~~~~~~~~~~~~~~~~~~~~~~~~~~~~~~~~~~~~~~~~~~~~~~~
\subsection{Data Quality Vetoes}

While ideally all sources of noise transients would be identified and noise sources mitigated through hardware corrections, in practice this is not always possible. Even when fixes to the instrument are made they may happen after weeks or months of polluted data being taking.
In that case we need to remove the artefacts from the already recorded data. For this purpose, Data quality (DQ) flags are produced, which mark second integer periods of time (called segments) when the GW detector output is to be considered suspect, because it may contain an excess of noise transients. Each search then assigns to the flags categories that dictate how the flag is to be used~\cite{CBCglitch}. These categories are rationalized as follows. 

\begin{itemize}
\item Category 1 flags veto instances when a serious problem with the detector occurred, or when the detectors is not taking data with near its nominal sensitivity. This data is removed prior to running the analysis. 
\item Category 2 flags veto times when a well understood problem with demonstrated coupling into the GW data channel occurred. These events are removed before the analysis determines the statistical significance of any candidate events.
\item Category 3 flags veto intervals when an incompletely understood or defined problem occurred.
These flags are used to provide a cleaner search background, with respect to category 2, and may also cast doubt on any candidate events.
\end{itemize} 

A number of DQ flags, based on auxiliary channels, are generated in near real-time (within a few minutes) by LIGO and Virgo. One goal of the data quality groups is to produce as many ``online" DQ flags as possible to allow rapid identification of problems, and rapid application of DQ flags for online searches and electromagnetic follow up of candidates. However so far the LIGO low latency online DQ flags remove only a small fraction of noise transients. 

There are also DQ flags automatically produced with a week latency by the aforementioned statistical ranking algorithms (hierarchical veto and UPV). These flags define sub-second time intervals that target individual noise transients. 

The most effective DQ flags are generated offline to target troublesome populations of noise transients. These are based on studies of populations of triggers in the GW outputs, and DQ flags are usually developed based on triggers, RMS, or thresholds on data from auxiliary channels that are found to be predictive of the observed noise. If the noise source in question is persistent, these flags are eventually made into online DQ flags. Figure~\ref{f:tailormadevetoes} illustrates the improvement in noise transient reduction seen for the S6 LIGO Hanford after applying targeted offline DQ flags. 

%------------------------------------
\subsubsection{What makes a suitable DQ flag}

DQ flag performance is estimated by applying the flag segments to the output of a search algorithm like Omega, run on a single detector. A standard quantity used to evaluate DQ flag performance is the ratio:
\begin{equation*}
\centering
%$\kappa$ = $\varepsilon$/$\tau$
\kappa = \varepsilon/\tau
\end{equation*}
where the efficiency $\varepsilon$ is the percentage of total triggers removed by the flag during the analyzed time period, and the dead time $\tau$ is the percentage of analyzed total time removed by the flag. For example, in an analyzed data segment containing 500 triggers in 1000 seconds, one might find that a flag vetoes 200 of the triggers (40\% efficiency), while removing 50 seconds of science data (5\% dead time), resulting in a $\kappa$ of 8. Judged with respect to other flags developed for gravitational-wave data this would qualify as an effective flag. If a DQ flag were to be constructed from randomly distributed time segments during this same period, and if the 500 triggers were also randomly distributed, we would expect the DQ flag to veto the same percentage of triggers as the dead time it incurs, so $\kappa$ would be around 1. The $\kappa$ ratio, and other considerations such as knowledge of the underlying physical coupling and the severity of the targeted glitch population for data analysis, are used to determine which category a given flag should belong to for each search.

%-----------------------------------------
\subsubsection{Testing for safety}
Before any DQ flags are used in a GW analysis they must be proven ``safe'', meaning that they do not systematically veto GW signals (at least not at a rate greater than expected by chance). To test the safety of a veto we use {\em hardware injections} which are simulated GW signals (over a wide range of waveforms and SNRs) injected into the data by applying forces on the interferometer mirrors to induce a differential arm motion that mimics the effect expected from GW signals ~\cite{Brown:Hardware:injections}. We then count how many of the hardware injections are vetoed by a given DQ flag. Data quality flags that veto hardware injections at a rate greater than that expected for randomly chosen flags are considered unsafe and are not suggested for use in the GW analysis (for more information see \cite{Hveto:2011, UPV:2010}). However, because the injections are few and not randomly distributed, DQ flags are sometimes incorrectly found to be unsafe using this statistical approach. To not unnecessarily throw out good flags, we do a further follow up, using the raw data from the auxiliary sensor and the detector output, of any DQ flags found to be statistically unsafe. For example, if a given flag is found statistically unsafe because it vetoes one weak hardware injection (by chance), but we find in the follow up that it does not veto a number of stronger injections of the same morphology, we may still use the flag.

%%%%%%%%%%%%%%%%%%%%%%%%%%%%%%%%%%%%%%%------------------------------------------------
\section{Results and Examples}
This section provides examples of DQ flags used for the analysis of data from Virgo+ and Enhanced LIGO, and presents the overall DQ flag performance results for the three interferometers.

\subsection{Virgo+}

\begin{figure}
\centering
\begin{tabular} {cc}
\includegraphics[width=0.7\textwidth]{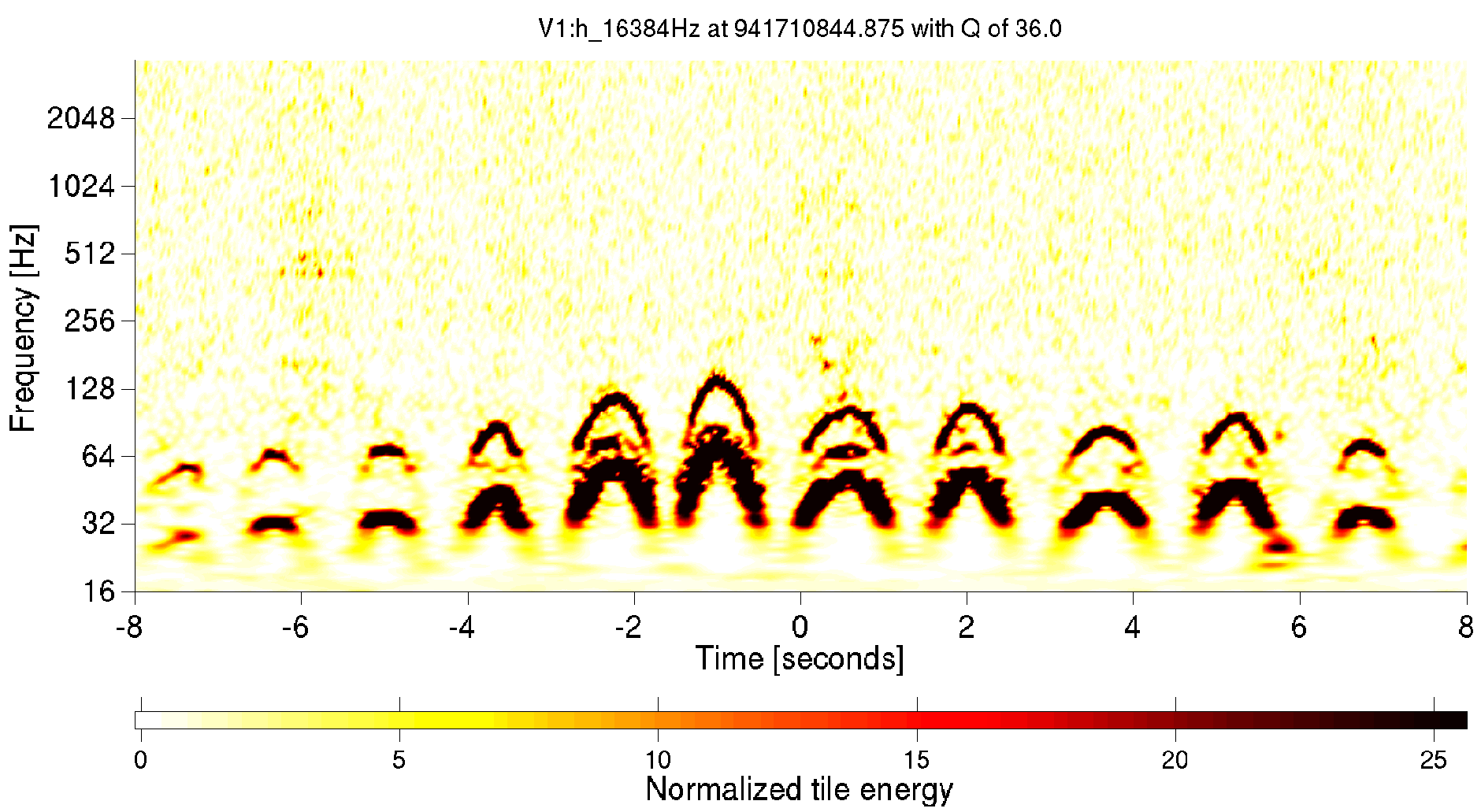} \\
\end{tabular}
\caption{\small{\textit{Time-frequency plot of the GW strain signal over a 16 second period around the time of the micro-seismic glitch}}}
\label{f:Virgoomegascan}
\end{figure}

%%%%% Virgo results %%%%%%%%

More than 100 DQ flags were produced online in Virgo, most of them used in categories 2 and 3 by GW searches. The types of issues that were flagged include saturation of photodiodes, glitches in the laser and control systems of the interferometer, and environmental disturbances such as seismic noise. 

\begin{figure}
\centering
\begin{tabular}{lcr}
\includegraphics[width=0.5\textwidth]{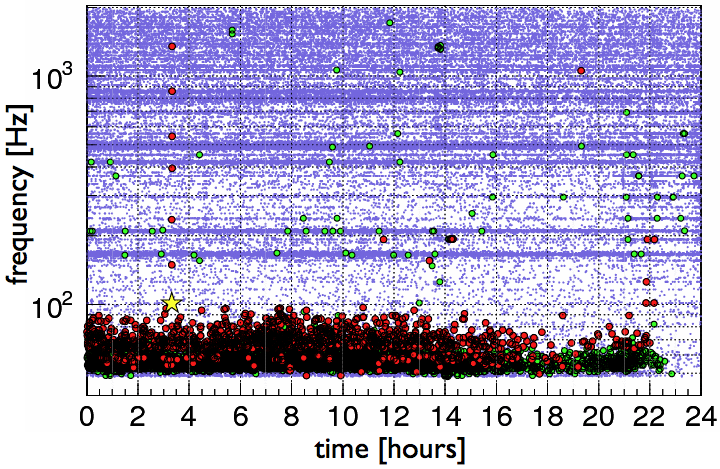} &
\includegraphics[width=0.5\textwidth]{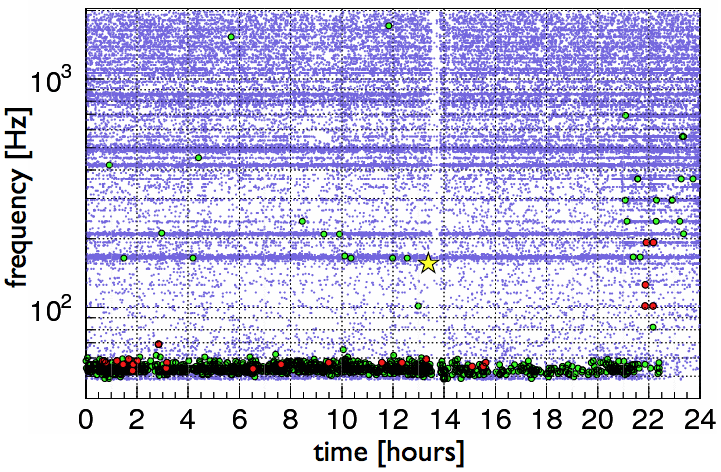}  \\
\end{tabular}
\caption{\small{\textit{Omega triggers in time-frequency before and after the event based
micro-seismic DQ flag has been applied}
}}
\label{f:Virgo_glitchgrams}
\end{figure}

One of the more important cat2 flags targeted a problematic glitch population during VSR2, noise transients caused by light back-scattered by optical components that are in contact with the ground and thus move at low and {\em micro-seismic} frequencies (0.1-16Hz).
This non-linear noise is clearly identified in spectrograms as arches with a peak frequency proportional to the velocity of the scatterer motion (see figure~\ref{f:Virgoomegascan}) \cite{Virgo:detchar:2010}. Virgo used signals recorded by seismometers and accelerometers on site to derive data quality flags for times when the local ground motion was large. These flags are effective at removing noise triggers from the detector output signal, as depicted in Figure~\ref{f:Virgo_glitchgrams}. One shortcoming of these flags is that they incur a high dead time during bad weather conditions. The overall performance of these seismic-based flags over the VSR2 run was an efficiency of 26.6\% on Omega triggers with $\mbox{SNR}> 8$ and a dead time of 1.7\%. The effects were greater at low frequency. For detector output triggers with central frequency below 100 Hz, the population containing most of the micro-seismic glitches \cite{VirgoDQ:2011}, the efficiency was 29.5\%.  

\begin{table}
\parbox{.45\linewidth}{
\centering
\begin{tabular}{lcr}
\hline
  Categories: & cat1+2 & cat1+2+3 \\
  Efficiencies:  & 54.2\% & 80.8\% \\
  dead times: & 6.2\% & 11.6\%  \\
\hline
\end{tabular}
\caption{VSR2 DQ flags performance for trigger SNR $>$8}
}\label{tab:virgoeffdt}
\hfill
\parbox{.45\linewidth}{
\centering
\begin{tabular}{lcr}
\hline
  Categories: & cat1+2 & cat1+2+3 \\
  Efficiencies:  & 19.3\% & 27.2\% \\
  dead times: & 4.2\% & 8.8\%  \\
  \hline
\end{tabular}
\caption{VSR3 DQ flags performance for trigger SNR$>$8}
}
\end{table}

 %%%
 
\begin{figure}
\centering
\begin{tabular} {cc}
\includegraphics[width=0.6\textwidth]{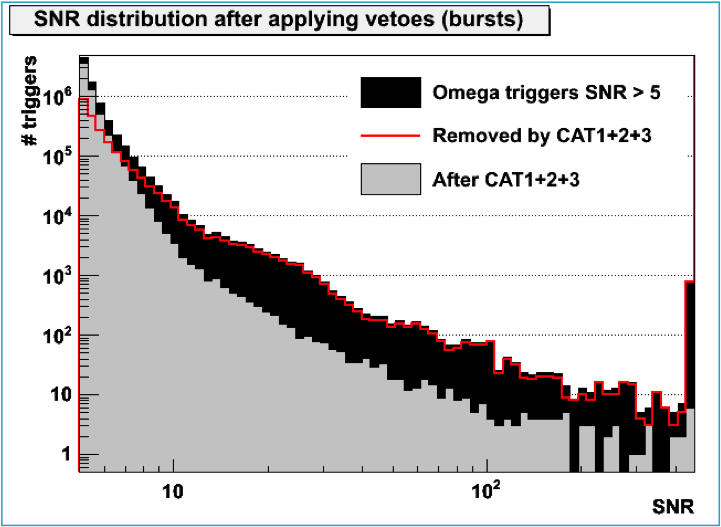}  \\
\end{tabular}
\caption{\small{\textit{VSR2 Omega trigger SNR distribution before and after DQ flags have been applied.
}}}
\label{f:VSR2}
\end{figure}

The cumulative effect on efficiency and dead time of the Virgo DQ flags on Virgo data during VSR2 and VSR3 is summarized in Table~\ref{tab:virgoeffdt}. The results for VSR2 were good - most of the loudest glitches in VSR2 were vetoed. Figure~\ref{f:VSR2} shows the VSR2 Omega trigger SNR distribution before and after cat1, 2,and 3 DQ flags have been applied. The moderately worse VSR3 performance (table 2) is mainly due to a high rate of noise transients in the first weeks of the run, caused by a high level of diffused light at the output of the Virgo interferometer, whose origin was an radius of curvature asymetry in the interferometer's arms end mirrors. 

\subsection{Enhanced LIGO} %%%%%%%%%%%%% eLIGO %%%%%%%%%%%%%

%%%% Seisveto example %%%%

Similar to Virgo, a large number of data quality flags were produced during LIGO's sixth science run. These included flags for photodiode saturations, magnetic and seismic disturbances, high wind, control system glitches, incorrectly calibrated data, and increased noise near the boundaries of science operation. 

One of the more successful DQ flags during Enhanced LIGO was the Seisveto flag applied to Hanford science run 6 (S6) data ~\cite{seisveto}. This flag was derived from seismometer signals on the Hanford site. The Omega burst detection pipeline was tuned to increase its sensitivity at low frequencies and was then run over data from several seismometers, producing a list of time segments corresponding to large low-frequency (below 32\,Hz) transient seismic noise. These segments were shown to have a high degree of time overlap with higher frequency transients in the LIGO Hanford detector output. The non-linear mechanism responsible for up-converting the seismic noise into the detector band is not fully understood.

The Seisveto flag produced the highest efficiency of any single flag during S6, with a 4.5\% dead time, 28\% efficiency, and a $\kappa$ of 6.2 on LIGO Hanford data triggers with an SNR $>$20 during the last epoch in S6 (known as ``S6D" and lasting about 4 months). However, it was only applied as category 3 to the unmodeled search because of its high dead time.

\begin{figure}
\centering
\begin{tabular} {ll}
Livingston & \\
\includegraphics[width=0.5\textwidth]{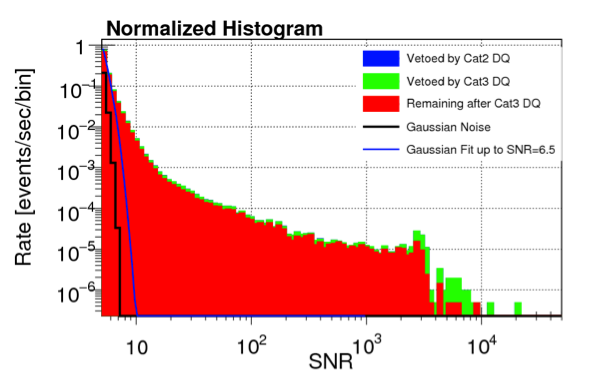}  &
\includegraphics[width=0.5\textwidth]{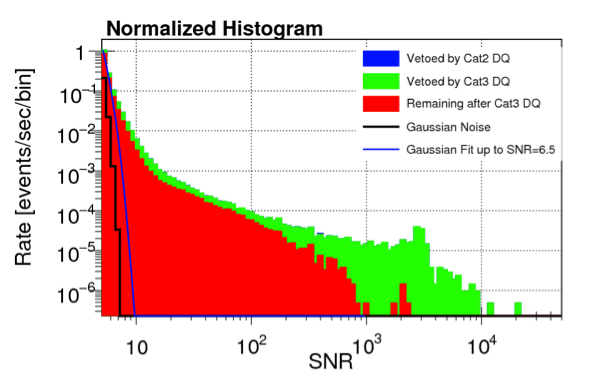} \\
Hanford & \\
\includegraphics[width=0.5\textwidth]{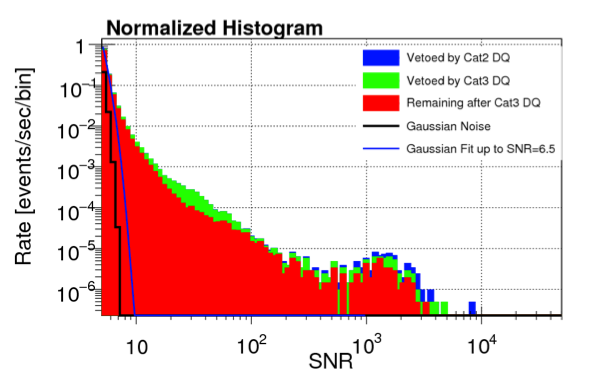}  &
\includegraphics[width=0.5\textwidth]{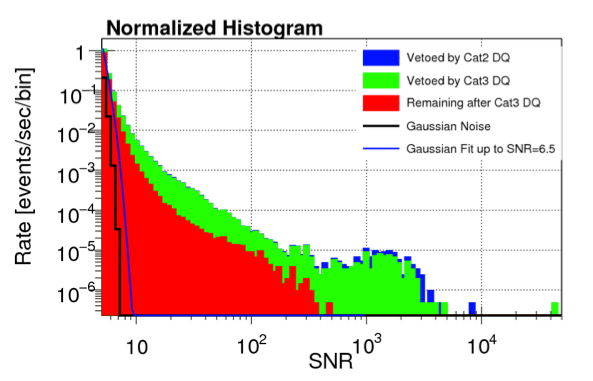} \\
(a) Online data quality vetoes applied  & (b) Online and targeted data quality vetoes applied \\
\end{tabular}
\caption{\small{\textit{LIGO Omega trigger SNR distribution during S6D. The difference between columns (a) and (b)
illustrates the effectiveness of targeted data quality vetoes vs. online vetoes. Both columns (a) and (b)
exclude event based vetoes.
}}}
\label{f:tailormadevetoes}
\end{figure}

The Enhanced LIGO detectors demonstrated improved noise transient behavior later in S6, with S6D~\cite{LIGODQ:2011}, having the lowest rate and SNR of the transients. Figures~\ref{f:tailormadevetoes} and~\ref{f:cWBhist}, show the trigger rates for S6A and S6D respectively. Figure~\ref{f:tailormadevetoes} also demonstrates the poorer performance of online vetoes for LIGO with respect to offline vetoes. The overall efficiency and dead time during S6Dare shown in Tables 3 and 4. The efficiency performance was quite low, however the flags performed far better on high SNR triggers. Category 3 efficiencies approached 100\%  for triggers of SNR $>$1000 for both of the LIGO detectors. The category 2 and 3 flags removed most of the high SNR trigger population, which was problematic for the searches (see column b of figure~\ref{f:tailormadevetoes}).

\begin{table}
\parbox{.45\linewidth}{
\centering
\begin{tabular}{lcr}
\hline
  Categories: & cat1+2 & cat1+2+3 \\
  Efficiencies:  & 3.8\% & 39.8\% \\
  dead times: & 0.3\% & 16\%  \\
\hline
\end{tabular}
\caption{LHO S6D DQ flags performance for trigger SNR $>$8 as defined by the burst (unmodeled) analysis}
}
\hfill
\parbox{.45\linewidth}{
\centering
\begin{tabular}{lcr}
\hline
  Categories: & cat1+2 & cat1+2+3 \\
  Efficiencies:  & 1.7\% & 29.4\% \\
  dead times: & 0.7\% & 14.3\%  \\
  \hline
\end{tabular}
\caption{LLO S6D DQ flags performance for trigger SNR $>$8 as defined by the burst (unmodeled) analysis}
}
\label{f:LIGO_S6D_results}
\end{table}

%~~~~~~~~~~~~~~~~~~~~~~~~~~~~~~~~~~~~~~~~~~~~~~~~~~
\section{Data Quality in the Advanced Detector Era}
The LIGO and Virgo collaborations have made progress in noise transient abatement during the recent
science runs. LIGO and Virgo's online veto production provided useful results in the latest science runs, but a large fraction of low or medium SNR (SNR$<$10) glitch populations remained unexplained and/or unvetoed, motivating the development of new vetoing tools in preparation for the next era of detectors.

LIGO is currently being upgraded to Advanced LIGO, and Virgo is just starting its Advanced configuration installation. The Advanced detectors are expected to be quantum-noise-limited interferometers, with an order of magnitude improvement in strain sensitivity. Studies predict that Advanced LIGO and Advanced Virgo should routinely detect gravitational waves from astrophysical sources~\cite{Abadie:2010cf}. Advanced era data quality studies will aim to provide low-latency and reliable information to help extract the maximum GW range and detection likelihood from these searches. This means to optimize the efficiency over dead time ratios of DQ flags, develop new vetoing tools, and continue to explore and implement new techniques for targeting noise transients.

%~~~~~~~~~~~~~~~~~~~~~~~~~~~~~~~~~~~~~~~~~~~~~~~~~~
\begin{ack}
The authors gratefully acknowledge the support of the United States
National Science Foundation for the construction and operation of the
LIGO Laboratory, the Science and Technology Facilities Council of the
United Kingdom, the Max-Planck-Society, and the State of
Niedersachsen/Germany for support of the construction and operation of
the GEO600 detector, and the Italian Istituto Nazionale di Fisica
Nucleare and the French Centre National de la Recherche Scientifique
for the construction and operation of the Virgo detector. The authors
also gratefully acknowledge the support of the research by these
agencies and by the Australian Research Council, 
the International Science Linkages program of the Commonwealth of Australia,
the Council of Scientific and Industrial Research of India, 
the Istituto Nazionale di Fisica Nucleare of Italy, 
%%---- modified Feb2012:
% the Spanish Ministerio de Educaci\'on y Ciencia, 
the Spanish Ministerio de Econom\'a y Competitividad,
%%----------------------
the Conselleria d'Economia Hisenda i Innovaci\'o of the
Govern de les Illes Balears, the Foundation for Fundamental Research
on Matter supported by the Netherlands Organisation for Scientific Research, 
the Polish Ministry of Science and Higher Education, the FOCUS
Programme of Foundation for Polish Science,
the Royal Society, the Scottish Funding Council, the
Scottish Universities Physics Alliance, The National Aeronautics and
Space Administration, the Carnegie Trust, the Leverhulme Trust, the
David and Lucile Packard Foundation, the Research Corporation, and
the Alfred P. Sloan Foundation.
\end{ack}

%~~~~~~~~~~~~~~~~~~~~~~~~~~~~~~~~~~~~~~~~~~~~~~~~~~
%\newpage
\section*{References}
\bibliographystyle{iopart-num}
\bibliography{references}

\end{document}